# Antiferromagnetic ordering in heavy-fermion system $Ce_2Au_2Cd$

S. Rayaprol and R. Pöttgen*

*Institut für Anorganische und Analytische Chemie, Westfälische Wilhelms-Universität*

*Münster, Corrensstrasse 30, D-48149, Germany*

*Email : pottgen@uni-muenster.de

## *Abstract*

*$La_2Au_2Cd$ and $Ce_2Au_2Cd$ were prepared from the elements by reactions in sealed tantalum tubes in a water-cooled sample chamber of an induction furnace. These intermetallics crystallize with the tetragonal $Mo_2FeB_2$ type, space group P4/mbm. While $La_2Au_2Cd$ is Pauli paramagnetic, $Ce_2Au_2Cd$ shows Curie-Weiss behaviour above 100 K with an experimental magnetic moment of 2.41(2) $\mu_B$/Ce atom, indicating trivalent cerium. Antiferromagnetic ordering is detected for $Ce_2Au_2Cd$ at 5.01(2) K and magnetization measurements reveal a metamagnetic transition at 3 K at a critical field of around 20 kOe with a saturation moment of 1.50(2) $\mu_B$/Ce atom at 80 kOe. The low-temperature heat capacity properties characterize $Ce_2Au_2Cd$ as a heavy fermion material with an electronic specific heat coefficient (γ) = 807(5) mJ/mol $K^2$ as compared to $La_2Au_2Cd$ with γ = 6(5) mJ/mol $K^2$.*

## I. INTRODUCTION

The $Ce_2T_2In$ indides and $Ce_2T_2Sn$ stannides (T = late transition metal) with ordered $U_3Si_2$ type structure have intensively been investigated in recent years with respect to their greatly varying magnetic and electrical properties.[1 and ref. therein] $Ce_2Ni_2In$ and $Ce_2Rh_2In$ are intermediate-valence systems. $Ce_2Pt_2In$ is a strongly temperature dependent paramagnet, and $Ce_2Cu_2In$, $Ce_2Pd_2In$, and $Ce_2Au_2In$ order magnetically at 5.5, 4.3, and 3.2 K, respectively.[2] $Ce_2Pt_2In$ shows strong Kondo type interactions and a non-magnetic heavy fermion ground state [2,3] with an electronic specific heat coefficient γ = 500 mJ/mol $K^2$.[4] Interesting behaviour was observed for the palladium compounds $Ce_2Pd_2Sn$ and $Ce_2Pd_2In$ which show small ranges of homogeneity $Ce_2Pd_{2+x}Sn_{1-x}$ and $Ce_2Pd_{2+x}In_{1-x}$. The increase in palladium content leads to a decrease of the magnetic ordering temperature.[5, 6] The cerium atoms in $Ce_2Pt_2Sn$ are trivalent and this stannide shows two magnetic transitions at 2.5 and 6.5 K.[7]

In parallel to the stannides, also some plumbides $Ce_2T_2Pb$ ($T$ = Rh, Pd, Pt, Au)[5, 8–10] have been synthesized. $Ce_2Pd_2Pb$[5] is a 6.2 K antiferromagnet and $Ce_2Rh_2Pb$[10] behaves like a normal metal.

Recently, the first cadmium containing compounds have been reported,[8, 11–14] where the $X$ site is completely occupied by cadmium. This substitution leaves opportunities for varying the properties, since cadmium reduces the valence electron count per formula unit by two with respect to the stannides and plumbides. Two highly interesting compounds are $Ce_2Ni_{1.88}Cd$ [11] and $Ce_2Rh_{1.86}Cd$ [13] which both show intermediate valence of cerium. In view of these promising results we have started a more systematic study of the $Ce_2T_2Cd$ intermetallics and their non-magnetic counterparts $La_2T_2Cd$. Herein we report on the magnetic properties and a specific heat study on $La_2Au_2Cd$ and $Ce_2Au_2Cd$.

## II. EXPERIMENTAL DETAILS

Starting materials for the synthesis of $La_2Au_2Cd$ and $Ce_2Au_2Cd$ were lanthanum and cerium ingots (Johnson Matthey), a gold bar (Heraeus), and a cadmium rod (Johnson Matthey, Ø 8 mm), all with stated purities better than 99.9%. Because of the low boiling point of cadmium (1083 K),[15] the samples were prepared in sealed high-melting tubes. Pieces of the rare earth elements, the gold bar, and the cadmium rod were weighed in the ideal 2:2:1 atomic ratio and arc-welded[16] in small tantalum tubes (ca. 1 cm$^3$ tube volume) under an argon pressure of ca. 600 mbar. The argon was purified over silica gel, molecular sieves, and titanium sponge (900 K). The sealed tantalum tubes were then placed in a water-cooled quartz sample chamber of a high-frequency furnace (Hüttinger Elektronik, Frieburg, Type TIG 1.5/300) under flowing argon.[17] The elements were brought to reaction through inductive annealing at ca. 1500 K for about one minute and the products were subsequently annealed at ca. 900 K for another two hours. The products could easily be separated from the tantalum tubes via mechanical fragmentation. No reaction of the samples with the crucible material was detected. For further details concerning the sample preparation we refer to Ref. 12.

The purity of the samples was checked though Guiner powder patterns using $CuK_{\alpha1}$ radiation and α-quartz ($a$ = 491.30, $c$ = 540.46 pm) as an internal standard. The magnetic and calorimetric measurements were performed on a Quantum Design PPMS with ACMS and specific heat options in the temperature range of 3 to 300 K, with magnetic flux densities up to 80 kOe. For magnetization measurements, the samples were enclosed in thin-walled gelatin capsules. The specific heat measurements were carried out with a relaxation technique. The samples were mounted on the sample holder with *Apizeon N grease.*

## III. RESULTS AND DISCUSSIONS

*Structure*

$La_2Au_2Cd$ and $Ce_2Au_2Cd$ crystallize with the tetragonal $Mo_2FeB_2$ type structure,[18] space group *P4/mbm*. In Figure 1, we show the XRD patterns for both compounds along with the theoretically expected peak positions (shown by vertical ticks). Both compounds were obtained as X-ray pure materials. The cell constants were calculated by least-square refinements of the powder data. The correct indexing was ensured through intensity calculations.[19] Cell constants and volumes as given in Fig. 1 agree with in experimental error with the values of $a$ = 809.2(1) pm, $c$ = 400.27(9) pm for $La_2Au_2Cd$ and $a$ = 804.93(7) pm, $c$ = 393.36(6) pm for $Ce_2Au_2Cd$ previously reported in Refs. [8, 12].

*Magnetic measurements*

Figure 2(a) shows the *dc* magnetic susceptibility measured in a stable field of 10 kOe for $Ce_2Au_2Cd$. The samples were zero field cooled to the lowest temperature, measurement was performed while warming the sample up to room temperature. $La_2Au_2Cd$ does not show any magnetic behavior and is paramagnetic to the lowest temperature measured. The room temperature susceptibility of 2.93(3) x $10^{-4}$ emu/mol is consistent with Pauli paramagnetism.

The χ(T) curve of $Ce_2Au_2Cd$ exhibits an definite peak at T ~ 5K indicating antiferromagnetic ordering. The inverse susceptibility ($\chi^{-1}$) follows Curie-Weiss behavior above 100 K. However, the deviation of $\chi^{-1}$ below 100 K, can be attributed to a combination of crystal field effects and magnetic ordering. From the CW fit of the linear region in $\chi^{-1}$ in the temperature region 100 < T(K) < 300, the paramagnetic Curie temperature ($\theta_p$) and the effective Bohr magneton number ($\mu_{eff}$) for $Ce_2Au_2Cd$ is -3.3(2) K and 2.41(2) $\mu_B$/mol-Ce respectively. The $\mu_{eff}$ obtained experimentally is in close agreement with value of the free $Ce^{3+}$ ion (2.54 $\mu_B$) indicating that unlike some of the $Ce_2T_2Cd$ compounds (T = Ni, Rh),[11,13] cerium is in trivalent state in $Ce_2Au_2Cd$. This is in excellent agreement with the course of the lattice parameters (lanthanoid contraction) for the series $RE_2Au_2Cd$, $RE$ = La, Ce, Pr, Nd and Sm, [12] where $Ce_2Au_2Cd$ shows no anomaly.

The negative sign of $\theta_p$ indicates the magnetic interactions are antiferromagnetic. There is no effect of H on the ordering temperature, while measuring $\chi(T)$. The insert in Fig 2a, shows $\chi(T)$ for $Ce_2Au_2Cd$ measured in H = 0.1, 1 and 10 kOe applied fields. It is interesting to observe that though the susceptibility depends upon the excitation field but the ordering temperature is not affected by it. In the ZFC (symbols)-FC(continuous lines) $\chi(T)$ for H = 0.1 and 1 kOe, FC curve follows the ZFC curve and there is no bifurcation between them.

The real part of ac susceptibility ($\chi'$), shown in Figure 3, exhibits a prominent peak at $T_N$ with no frequency dependence. The imaginary part ($\chi''$) is essentially featureless with a broad feature seen only at higher frequencies, thus ruling out any spin-glass anomalies. There are no features in $\chi_{ac}$ measured for second and third harmonics (and hence not shown here), thus ruling out any ferromagnetic impurities also. These observations clearly establish long-range magnetic ordering of the antiferromagnetic type at 5 K in $Ce_2Au_2Cd$.

The magnetization as a function of applied magnetic field at different temperatures spanning $T_N$ are shown in Figure 4. The magnetization at temperatures 300 and 100 K (>> $T_N$) varies linearly with the application of field. However for T = 10 K, the magnetization increases linearly up to a field of 60 kOe and deviates slightly at higher fields. For M(H) at T = 4.5 K (i.e., just below $T_N$), M varies sluggishly with H without saturating up to 80 kOe. The M(H) at 3 K, (T < $T_N$), exhibits a metamagnetic transition starting around 20 kOe and increases with H without saturating, up to the highest field measured. It may be recalled that an isostructural compound, $Ce_2Au_2In$ also exhibits such a step-like increase in magnetization[2,3] at T < $T_N$, but the metamagnetic transition already takes place around 5 kOe. However unlike $Ce_2Au_2In$, magnetization for $Ce_2Au_2Cd$ at 3 K does not saturate. The moment value for $Ce_2Au_2Cd$ at 80 kOe and 3 K is 1.53(3) $\mu_B$/Ce atom, smaller than the maximum value of 2.14 $\mu_B$/Ce atom according to *g* x *J*. An even smaller saturation magnetization of only 0.97 $\mu_B$/Ce atom was observed for $Ce_2Au_2In$ at 1.7 K and 50 kOe.[2]

*Specific heat studies*

Heat capacity, C(T), for $Ce_2Au_2Cd$ and its isostructural non-magnetic counterpart $La_2Au_2Cd$ were measured by relaxation method using the PPMS heat capacity option in the temperature range 3–100 K on a puck calibrated for temperature and magnetic field. We have also measured the heat capacity for $Ce_2Au_2Cd$ in presence of applied dc fields of 10, 20, 30 and 50 kOe.

Figure 5 (a and b) shows C(T) and $CT^{-1}$ vs. T respectively for both $Ce_2Au_2Cd$ and $La_2Au_2Cd$. The heat capacity for the lanthanum compound varies with temperature and essentially shows contributions from the lattice. However, the cerium compound exhibits a sharp peak in C(T) at 5 K undergoing magnetic transition. This is also a confirmation for the ordering temperature from magnetization measurements.

The magnetic part of heat capacity ($C_{mag}$) was deduced after subtracting the lattice part (i.e., the heat capacity of non-magnetic $La_2Au_2Cd$) from the total heat capacity of $Ce_2Au_2Cd$. The peak at 5 K can be clearly seen in $C_{mag}$ also. It is interesting to observe that from the bottom of the peak (~ 5.5 K) up to 29 K (which is equal to $\Theta_D$) there is hardly any change in $C_{mag}$ but beyond 29 K, $C_{mag}$ increases rapidly with increasing temperature.

The plot of $C_{mag}/T$ vs. $T^2$ is linear below $T_N$ and is shown in the insert of Figure 6a. The $T^3$ behavior of $C_{mag}$ in the ordered region is a typical feature of heavy fermions exhibiting antiferromagnetic magnetic ordering.[20, 21] It is interesting to observe that the value of the coefficient of the electronic specific heat 'γ' (Sommerfeld coefficient) obtained from the linear fit below $T_N$, is 807(5) mJ/mol $K^2$. The values of the coefficient of thermal expansion (β) and the Debye temperature ($\Theta_D$) are 80 mJ/mol $K^4$ and 29 K, respectively. For comparison, the isostructural counterpart of the title compound is $Ce_2Au_2In$. It exhibits antiferromagnetic ordering at around 3 K, but has a γ of 37 mJ/mol $K^2$.[3] Among other isostructural indides, the highest γ = 500 mJ/mol $K^2$ is observed for non-magnetic heavy fermion system $Ce_2Pt_2In$.[4] To the best of our knowledge, $Ce_2Au_2Cd$ is the first compound among cadmium based 221 intermetallics exhibiting such a high value of γ and hence be called a heavy fermion compound exhibiting antiferromagnetic ordering. At this point it is worthwhile to note, that in the family of actinide (*An*) intermetallics *An$_2$T$_2$In*, U$_2$Pt$_2$In shows an even larger γ value of 850 mJ/mol K$^2$, while U$_2$Pt$_2$Sn (334 mJ/mol K$^2$) and U$_2$Pd$_2$In (393 mJ/mol K$^2$) have slightly smaller Sommerfeld coefficients.[22,23] In the context of cerium intermetallics Ce$_2$Au$_2$Cd can be discussed in line with the equiatomic compounds CePdIn (700 mJ/mol K$^2$) and CePtIn (> 500 mJ/mol K$^2$).[24-26]

In Fig. 6 (b) we show the variation of magnetic entropy (S$_{mag}$) as a function of temperature. At the ordering temperature, S$_{mag}$ reaches the value of ~7.93 J/mol K, which is about 75% of the expected Rln2 value. The entropy reaches the 100% value of Rln2 at T~ 29 K (which incidentally is equal to Θ$_D$). Beyond Θ$_D$, S$_{mag}$ increases linearly with T.

We have also studied the effect of the external magnetic field on C(T). Figure 7 (a) and (b) shows C and C/T vs. T for Ce$_2$Au$_2$Cd measured in zero field and the applied dc fields of 10, 30 and 50 kOe. As expected for an antiferromagnet, the peak temperature (T$_p$) and magnitude of C, shifts to lower values with increasing field for Ce$_2$Au$_2$Cd. There is a small observable change in T$_p$ for H = 10 kOe. But with increasing the field to 30 kOe, T$_p$ shifts to lower temperature and exhibits two broad transitions. The magnitude of C however is further lowered. The appearance of a secondary peak at lower temperatures with application of 30 kOe field may imply field induced changes in the magnetic structure. At 50 kOe the peak is completely smeared out. The C$_H$(T) curves crosses each other just above the ordering temperature. A qualitative discussion about the crossing points in specific heat curves is given by Vollhardt et al.[27] Though the observation of crossing of C(T) curves for low values of H were initially made in normal-liquid $^3$He, they can also be seen in heavy fermion systems with and without Fermi liquid behavior. Hence it will be interesting to correlate the high γ value and crossing of C$_H$(T) curves in the vicinity of second-order magnetic transition for Ce$_2$Au$_2$Cd in the context of heavy fermion behavior.

Crossing of specific heat curves, as discussed above, can also be linked to the change in entropy with respect to the applied field. A close look at Fig. 7(b) shows that C/T is linear below T$_p$. We have calculated the total entropy of Ce$_2$Au$_2$Cd for measurements performed in H = 0, 10 and 30 kOe fields.

In figure 8 (a) we plot the variation of total entropy of the system ($S_{tot}$) with temperature and the effect of magnetic field on it. We have also plotted, for clarity, the change in total entropy ($\Delta S_{tot}$) with application of field, as a function of temperature in Fig. 8(b). The curve S1 shows $S_{10kOe}-S_0$, i.e., difference in entropy measured at 0 and 10 kOe. Similarly S2 shows the difference of entropy in 0 and 30 kOe. S1 changes sign of $\Delta S$ around $T_N$, and becomes negative. However S2 is quite interesting. It exhibits a peak like feature around 3.6(2)K and decreases up to $T_N$, then again rises at higher temperatures. Recalling the M(H) curve at 3K in this context, we have seen a metamagnetic transition around 30 kOe, which is clearly seen as a peak in the plot of *dM/dH* vs. H. Such field induced behavior indicates towards possible modification of the Fermi surface, also supported by the enhancement of γ value. We strongly feel that a detailed investigation of the (magneto) transport properties would be quite rewarding.

A phenomenological band-structure approach for understanding the magnetic ordering characteristics of ternary intermetallics containing Ce (*4f*) and U (*5f*) as magnetic ions has been given by Endstra et.al.[29] In order to consistently understand the magnetic behavior of $Ce_2Au_2Cd$, one needs to carry out a systematic study of $Ce_2T_2Cd$ series. More insight into the mechanism underlying the absence or presence of magnetic-ordering of these ternary intermetallics can be gained by applying proper experimental and theoretical approach. The band-structure approach proposed by Endstra et.al however explains the phenomenon within the Doniach phase diagram[29] of the Kondo lattice. The magnetic properties of $Ce_2Au_2Cd$ give indication of the Kondo effect. Since $\theta_p$ is comparable to $T_N$, the entropy reduction due to the Kondo screening competes with the entropy reduction due to antiferromagnetic transition of the hybridized states. However, contrary to the usual Doniach picture, it seems that the antiferromagnetic transition here is due to the formation of the spin-density waves (SDW), which destabilizes a part of the Fermi surface. Hence, in order to understand the structural, magnetic and physical properties of $Ce_2Au_2Cd$, an estimate of the *f-d* hybridization strength by theory and detailed photoemission spectroscopy results will be quite rewarding.

To conclude, heavy fermion behavior was observed for the 5 K antiferromagnetic $Ce_2Au_2Cd$ while $La_2Au_2Cd$ is a simple Pauli paramagnet. The Ce$_2$T$_2$Cd intermetallics are an interesting family of compounds with promising ground-state properties. Further investigations on these materials are currently in progress.

## ACKNOWLEDGMENTS

This work was supported by the Deutsche Forschungsgemeinschaft. We thank Dr. R-D. Hoffmann for helpful discussions. S.R. is indebted to the Alexander von Humboldt foundation for a research stipend.

**FIG.1** XRD patterns (Guinier technique, Cu K$_{\alpha 1}$ radiation) for La$_2$Au$_2$Cd and Ce$_2$Au$_2$Cd. The theoretically calculated peak positions are given as ticks in the lower part of each diagram.

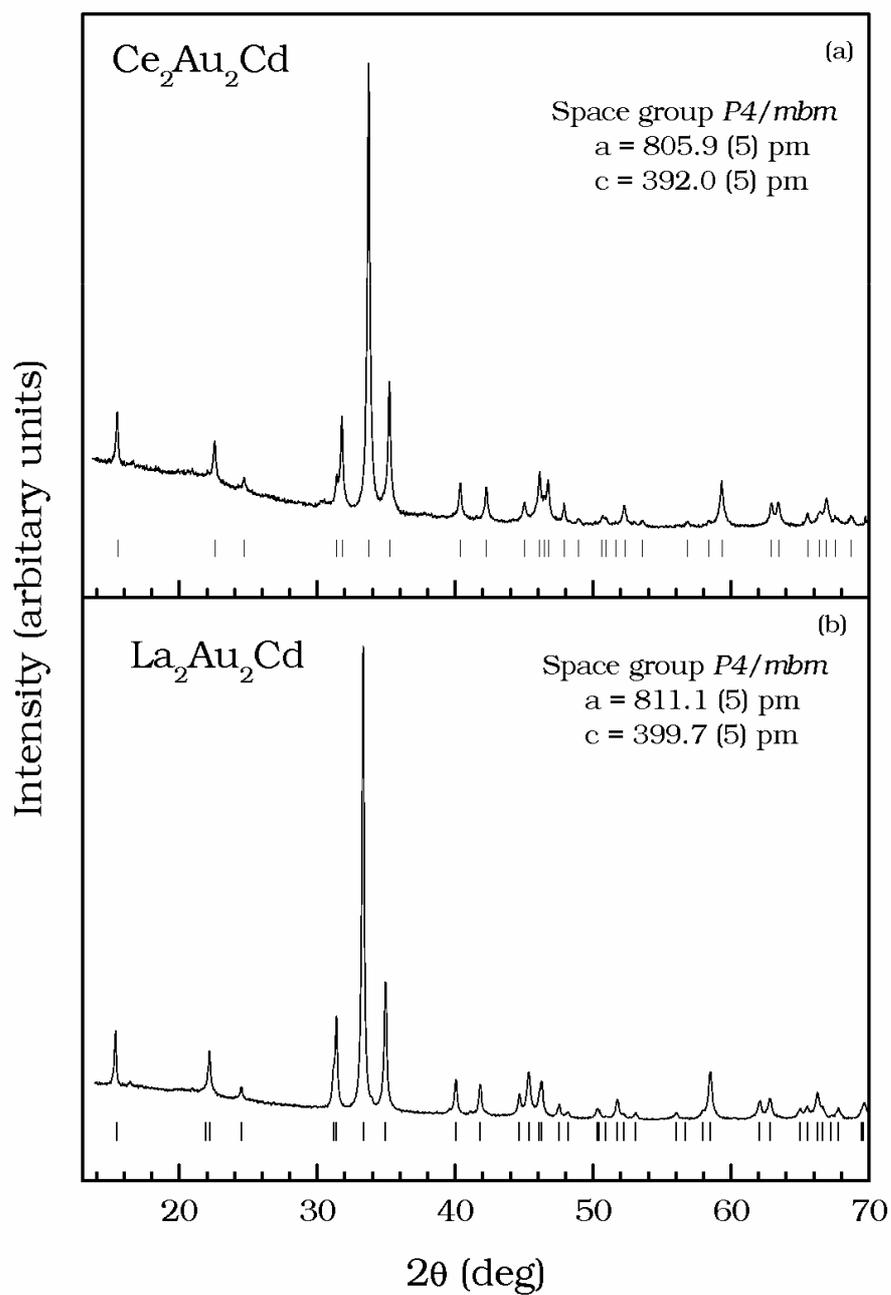

**FIG.2** (a) *dc* susceptibility ($\chi$) as a function of temperature (T) for $Ce_2Au_2Cd$, measured in applied field of 10 kOe. The insert shows $\chi(T)$ measured in different applied fields of H = 0.1, 1 and 10 kOe. The ZFC-FC $\chi(T)$ curves are distinguished by open circles (ZFC) and continuous line (FC). (b) $\chi^{-1}$ vs. T for $Ce_2Au_2Cd$ is shown by open circles. The continuous line is the fit from high temperature linear region. The arrow near the origin indicates $\theta_p$ extrapolated from the CW fit.

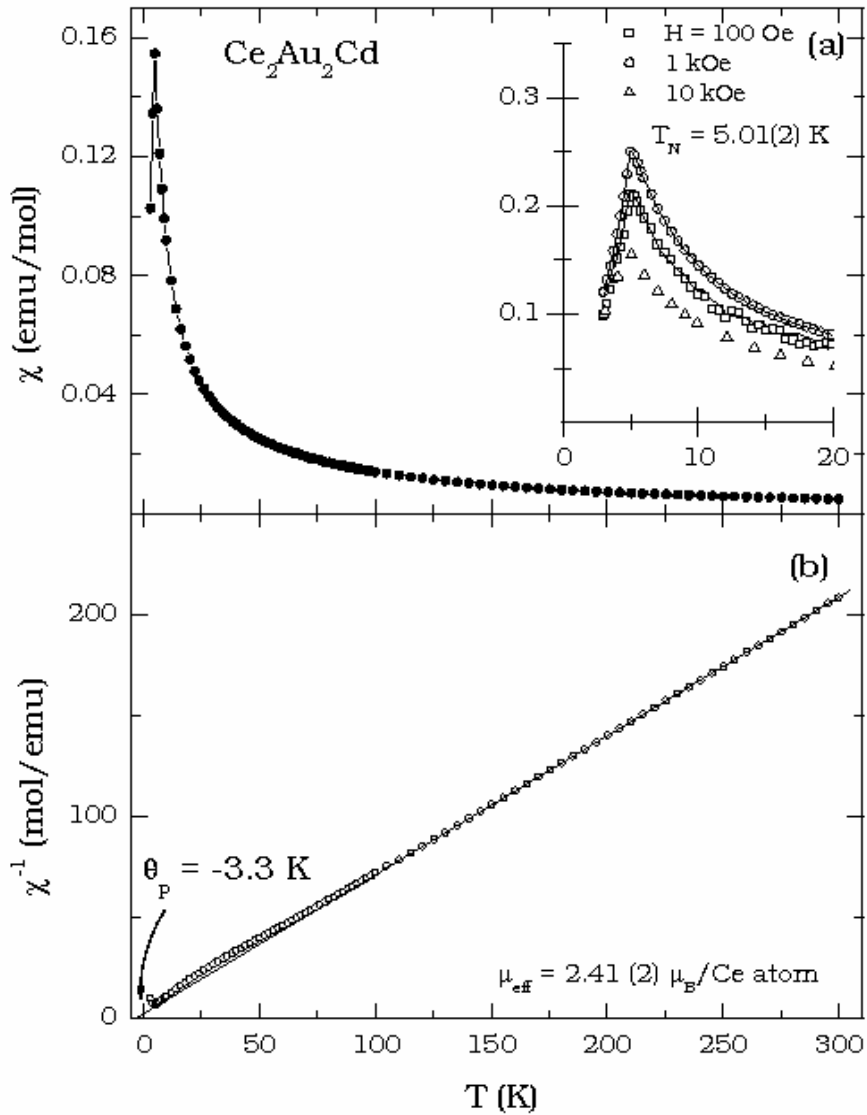

**FIG. 3** The real and imaginary part of the *ac* susceptibility ($\chi'$ and $\chi''$) vs. T respectively measured at various frequencies ($\nu$ = 111, 197, 341, 607, 1057, 1847, 3247, 5697 and 9999 Hz) and *ac* amplitude of 1 Oe for $Ce_2Au_2Cd$.

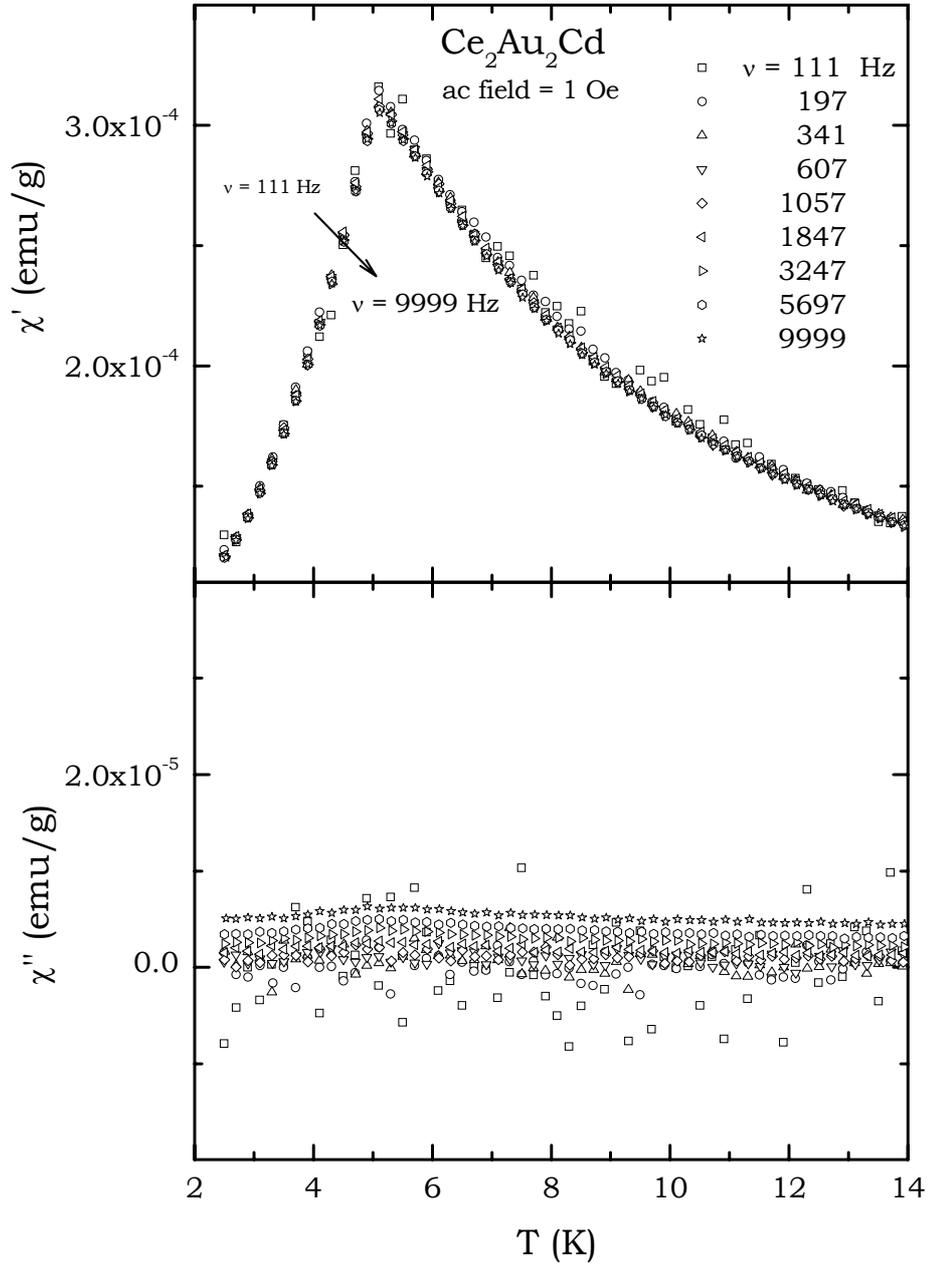

**FIG. 4** Magnetization (M) vs. applied field (H) at temperatures spanning $T_N$ for $Ce_2Au_2Cd$. The line passing through the data points shows a straight-line behavior of the magnetization below about 15 kOe.

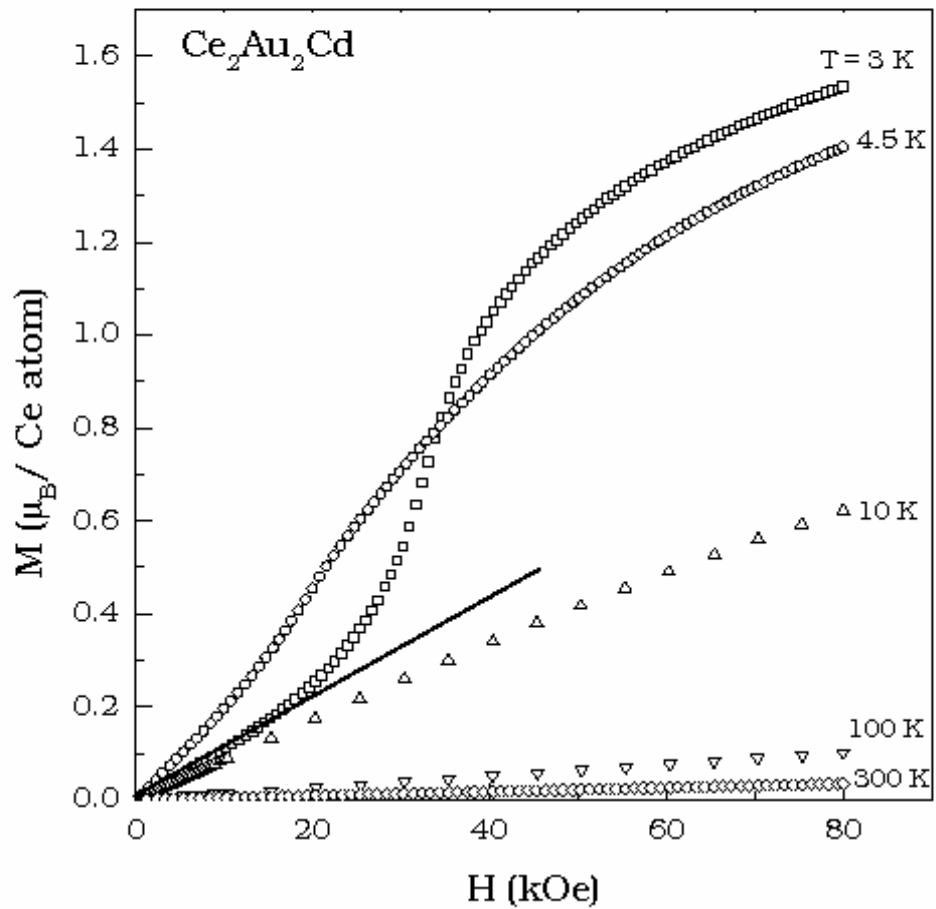

**FIG. 5** (a) Heat capacity (C) vs. T and (b) C/T vs. T for $RE_2Au_2Cd$ ($RE$ = Ce and La) shown by data points (Ce) and continuous line (La). The ordering temperature for $Ce_2Au_2Cd$ is indicated by the vertical arrow.

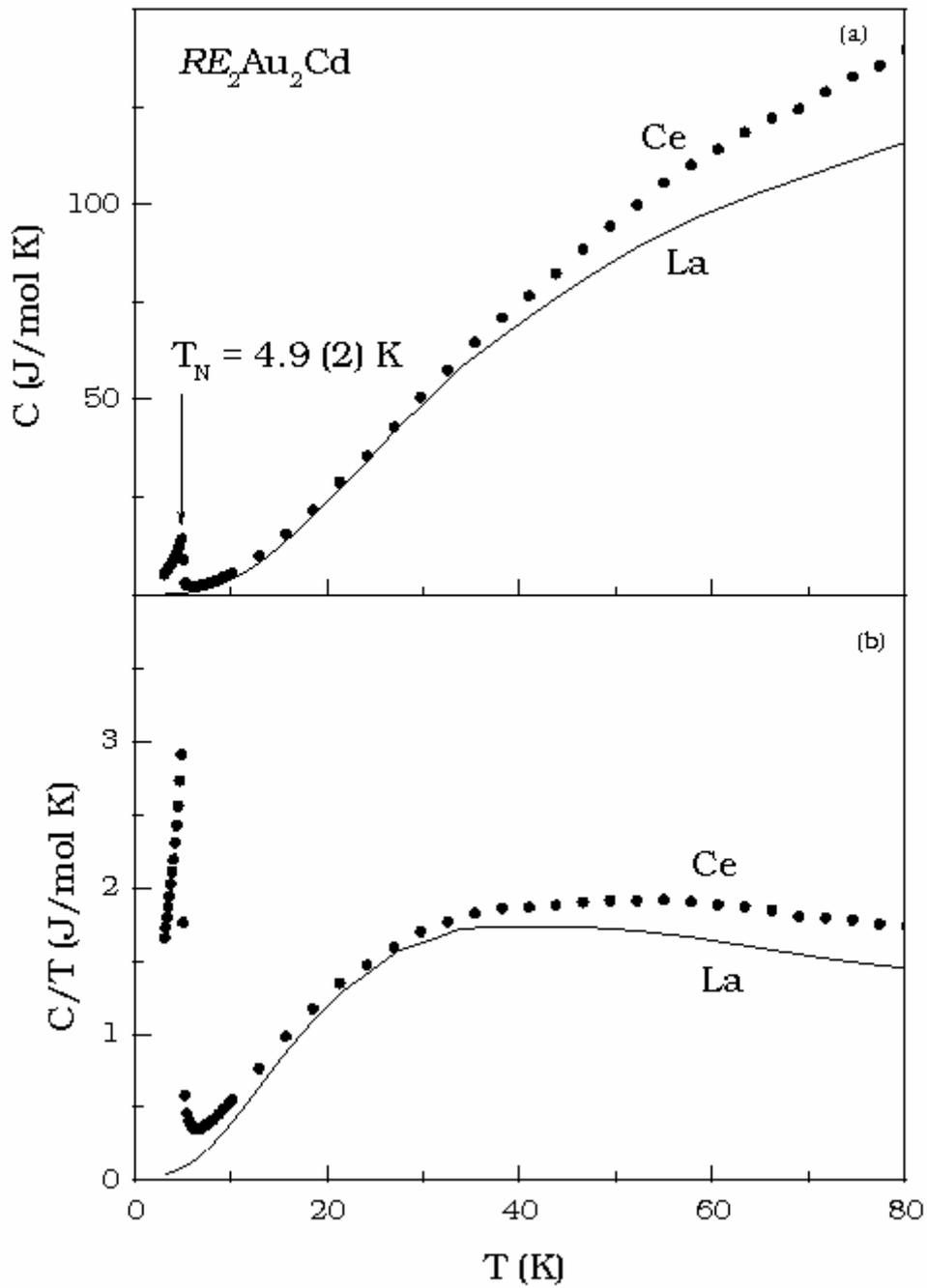

**FIG. 6** (a) The magnetic part of heat capacity ($C_{mag}$) for $Ce_2Au_2Cd$ obtained by subtracting the lattice part ($C_{La}$) from the total heat capacity of $Ce_2Au_2Cd$. The ordering temperature and Debye temperature are indicated by vertical arrows. The insert shows $C_{mag}/T$ vs. $T^2$ for $Ce_2Au_2Cd$ highlighting the $T^3$ behavior of $C_{mag}$ in the ordered state. The continuous line passing through the data points is a guide for the eyes only. (b) The magnetic entropy for $Ce_2Au_2Cd$. At the $T_N$, $S_{mag}$ reaches about 75% of Rln2 only (dotted line).

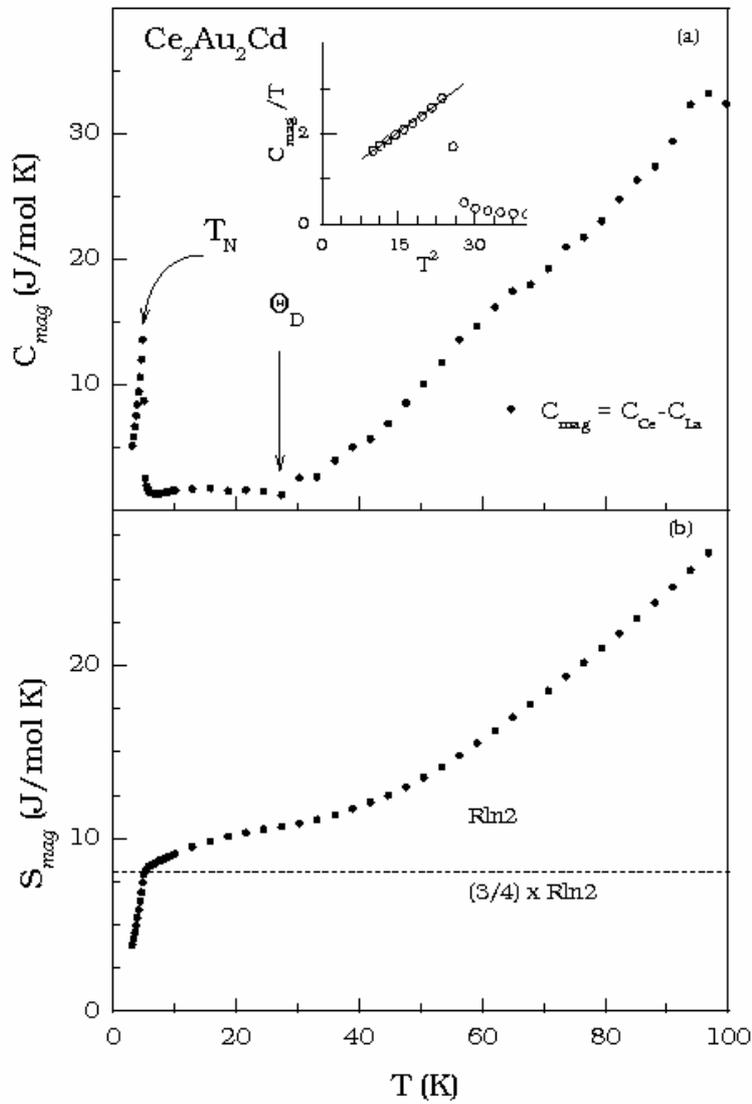

**FIG. 7** (a) Field dependence of heat capacity ($C_H$) and (b) $C_H/T$ vs. T, for $Ce_2Au_2Cd$. The crossing over point (see text) is indicated by vertical arrows.

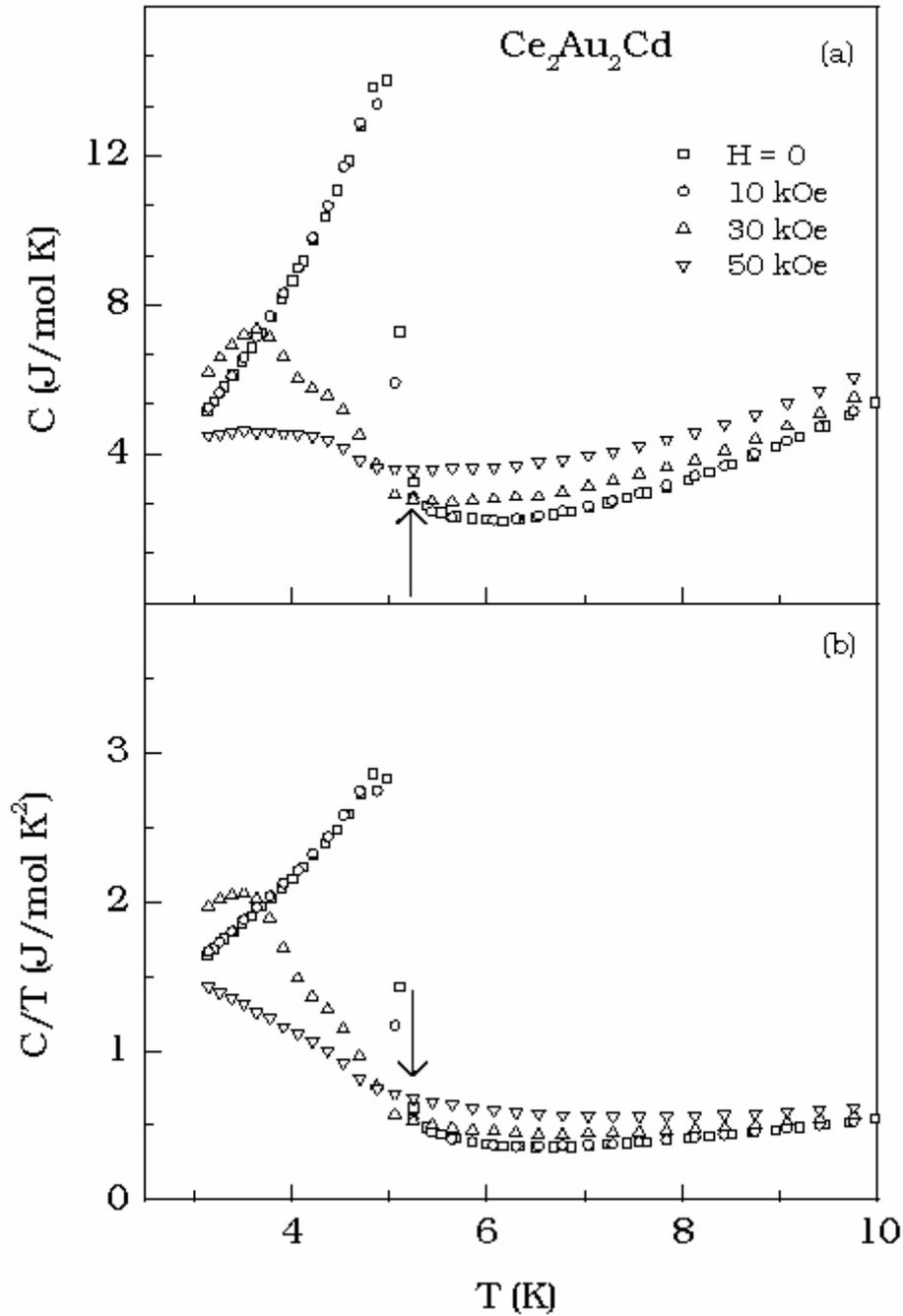

**FIG. 8** (a) Total entropy ($S_{tot}$) for $Ce_2Au_2Cd$ under applied fields. (b) The change in total entropy with the change in applied field, where $S1 = S_{10kOe} - S_0$ and $S2 = S_{30kOe}-S_0$.

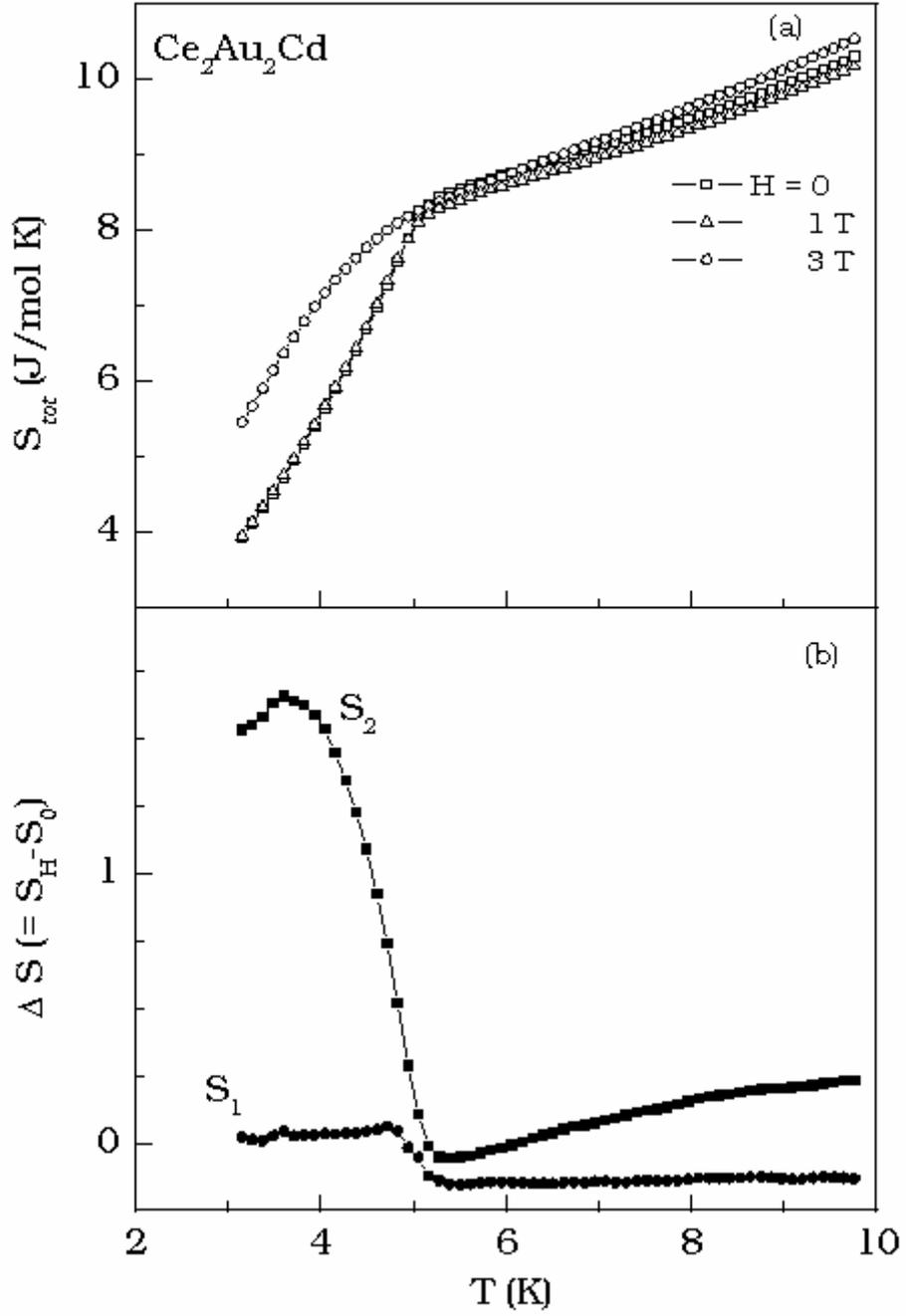